\documentclass[showpacs,twocolumn,floatfix,pra]{revtex4}
\usepackage{graphicx}
\usepackage{bm}
\begin{document}
\title{Relation between entanglement measures and Bell inequalities for three qubits}
\author{C. Emary and C. W. J. Beenakker}
\affiliation{Instituut-Lorentz, Universiteit Leiden, P.O. Box 9506, 2300 RA
Leiden, The Netherlands}
\date{November 2003}
\begin{abstract}
For two qubits in a pure state there exists a one-to-one relation 
between the entanglement measure (the concurrence ${\cal C}$) and 
the maximal violation ${\cal M}$ of a Bell inequality. No such 
relation exists for the three-qubit analogue of ${\cal C}$ 
(the tangle $\tau$), but we have found that numerical data is 
consistent with a simple set of upper and lower bounds for $\tau$ 
given ${\cal M}$. The bounds on $\tau$ become tighter with 
increasing ${\cal M}$, so they are of practical use. The Svetlichny 
form of the Bell inequality gives tighter bounds than the Mermin form. 
We show that the bounds can be tightened further if the tangle is 
replaced by an entanglement monotone that can identify both the
W state and the Greenberger-Horne-Zeilinger state.
\end{abstract}
\maketitle

Bell inequalities test for the quantum entanglement of a state by 
comparing the maximally measured value ${\cal M}$ of a certain 
correlator with the maximal value allowed by local realism 
\cite{Bel64}. For a pure state of two qubits, the Bell-CHSH 
(Clauser-Horne-Shimony-Holt \cite{Cla69}) parameter
${\cal M}=2\sqrt{1+{\cal C}^{2}}$ is directly related to the 
degree of entanglement (or concurrence) ${\cal C}\in[0,1]$ of the 
state \cite{Gis91}. This relation is useful because, on the one hand 
${\cal M}$ can be readily measured \cite{Asp81}, while on the other, 
${\cal C}$ can be readily calculated \cite{Woo98}. In this paper 
we investigate to what extent this relation has a three-qubit analogue.

The three-qubit analogue of the concurrence ${\cal C}$ is the 
tangle $\tau$, introduced by Coffman, Kundu, and Wootters \cite{Cof00}. 
It quantifies the irreducible tripartite entanglement, through the formula
\begin{equation}
  \tau={\cal C}_{A(BC)}^{2}-{\cal C}_{AB}^{2}-{\cal C}_{AC}^{2}.
  \label{taudef}
\end{equation}
The indices $A,B,C$ label the three qubits; the tangle is
invariant under permutation of these indices. 
The concurrence 
${\cal C}_{AB}$ refers to the mixed state of qubits $A$ and 
$B$ obtained after tracing out the degree of freedom of qubit 
$C$, and ${\cal C}_{AC}$ is defined similarly. The concurrence 
${\cal C}_{A(BC)}$ describes the entanglement of qubit $A$ with 
the joint state of qubits $B$ and $C$. The tangle $\tau\in[0,1]$ 
equals 0 if one of the qubits is separable from the other two. 
It equals 1 for the maximally entangled GHZ 
(Greenberger-Horne-Zeilinger \cite{Gre89}) state 
$|\psi\rangle_{\rm GHZ}=(|000\rangle+|111\rangle)/\sqrt{2}$.

The three-qubit analogue of the Bell-CHSH  inequality is due to 
Mermin \cite{Mer90}. There exists no analytical formula that
gives the maximal violation ${\cal M}_{\rm M}$ of the Mermin 
inequality for a given pure state of three qubits, but it is
not difficult to perform the maximization numerically. For 
special one-parameter states of the form 
$|\psi\rangle=\cos\alpha|000\rangle+\sin\alpha|111\rangle$, 
Scarani and Gisin \cite{Sca01} found an approximate 
(but highly accurate) relation 
${\cal M}_{\rm M}\approx\max(4\sqrt{\tau},2\sqrt{1-\tau})$ 
between $\tau=\sin^2 2\alpha$ and ${\cal M}_{\rm M}$. 

For more general states there is a range of values of $\tau$ with 
the same ${\cal M}_{\rm M}$. We have investigated this range 
numerically and found that 
the data is well described by a simple pair of upper and lower bounds 
for $\tau$  for any given ${\cal M}_{\rm M}$.
The bounds can be tightened in two ways: (1) By using an alternative 
form of the three-qubit Bell inequality, due to Svetlichny 
\cite{Sve87}; (2) By using an alternative measure $\sigma$ of 
tripartite entanglement that we introduce in this paper, defined by
\begin{equation}
  \sigma=\min\left(\frac{{\cal C}_{X(YZ)}^{2}
  +{\cal C}_{Y(XZ)}^{2}}{2}-{\cal C}_{XY}^{2}\right).\label{sigmadef}
\end{equation}
The minimisation is over the permutations $X,Y,Z$ of the qubits 
$A,B,C$. We find the following bounds on $\sigma$ for a given 
maximal violation ${\cal M}_{\rm S}$ of the Svetlichny inequality:
\begin{equation}
  |{\cal M}_{\rm S}^{2}/16-1|\alt\sigma
  \alt{\cal M}_{\rm S}^{2}/32.\label{MSsigmabounds}
\end{equation}
(We use the symbol $\alt$ instead of $\leq$ as these bounds are inferred 
from numerical data, rather than derived analytically.)

Both $\sigma$ and $\tau$ are entanglement monotones (meaning that 
they can not be increased on average by local operations and classical 
communication). Their essential difference is that $\sigma$ can 
detect tripartite entanglement of both the W and GHZ types, while 
it is known that $\tau$ can only detect GHZ type entanglement 
\cite{Dur00}. We recall that local operations on the W state 
$|\psi\rangle_{\rm W}=(|001\rangle+|010\rangle+|100\rangle)/\sqrt{3}$ 
and the GHZ state $|\psi\rangle_{\rm GHZ}$ generate two distinct 
classes of irreducibly entangled tripartite states. While 
$\tau=1=\sigma$ for $|\psi\rangle_{\rm GHZ}$, for $|\psi\rangle_{\rm W}$ 
only $\sigma=4/9$ is non-zero. In fact, $\sigma=0$ if and only if 
one of the qubits is separable from the other two (2--1 separability). 
This latter 
property distinguishes the entanglement measure introduced here 
from the one introduced by Meyer and Wallach \cite{Mey02}, which 
is also nonzero for 2--1 separable states.

After this introduction, we now present our findings in more detail.

Pure states of three qubits constitute a five-parameter family, 
with equivalence up to local unitary transformations. This family 
has the representation \cite{Aci00}
\begin{eqnarray}
  |\psi\rangle&=&\sqrt{\mu_{0}}|000\rangle
  +\sqrt{\mu_{1}}e^{i\phi}|100\rangle\nonumber\\
  &&\mbox{}+\sqrt{\mu_{2}}|101\rangle+\sqrt{\mu_{3}}
  |110\rangle+\sqrt{\mu_{4}}|111\rangle,\label{fiveparameters}
\end{eqnarray}
with $\mu_{i}\geq 0$, $\sum_{i}\mu_{i}=1$, and $0\leq\phi\leq\pi$. 
The labels $A,B$, and $C$ indicate the first, second, and third 
qubit, while $X,Y,Z$ refer to an arbitrary 
permutation of these labels.

The tangle (\ref{taudef}) is given by
\begin{equation}
 \tau=4\mu_{0}\mu_{4}.\label{taumu}
\end{equation}
The squared concurrences ${\cal C}_{X(YZ)}^{2}=4\,{\rm Det}\,\rho_{X}$ 
(with $\rho_{X}={\rm Tr}_{Y,Z}|\psi\rangle\langle \psi|$ 
the reduced density matrix) 
take the form
\begin{eqnarray}
  {\cal C}_{A(BC)}^{2}&=&4\mu_{0}(\mu_{2}+\mu_{3}+\mu_{4}),\label{CABCdef}\\
  {\cal C}_{B(AC)}^{2}&=&4\mu_{0}(\mu_{3}+\mu_{4})+4\Delta,\label{CBACdef}\\
  {\cal C}_{C(AB)}^{2}&=&4\mu_{0}(\mu_{2}+\mu_{4})+4\Delta,\label{CCABdef}
\end{eqnarray}
with the definition 
$\Delta=\mu_{1}\mu_{4}+\mu_{2}\mu_{3}-2(\mu_{1}\mu_{2}
\mu_{3}\mu_{4})^{1/2}\cos\phi$. 
Each of the four quantities (\ref{taumu})--(\ref{CCABdef}) 
is an entanglement monotone \cite{Dur00,Gin02}.

The quantity $\sigma$ defined in Eq.\ (\ref{sigmadef}) can 
equivalently be written as
\begin{eqnarray}
  \sigma&=&{\textstyle\frac{1}{2}}(\tau+\min{{\cal C}_{Z(XY)}^{2}})
  \nonumber\\
  &=&\tau+{\textstyle\frac{1}{2}}\min({\cal C}_{XZ}^{2}
  +{\cal C}_{YZ}^{2}),\label{sigma2def}
\end{eqnarray}
as follows from the
identity \cite{Sud01} 
$\tau={\cal C}_{X(YZ)}^{2}+{\cal C}_{Y(XZ)}^{2}
-{\cal C}_{Z(XY)}^{2}-2{\cal C}_{XY}^{2}$. 
One sees that $0\leq\tau\leq\sigma\leq 1$. Most importantly, 
since $\tau$ and $\min{\cal C}^2_{Z(XY)}$ are positive 
entanglement monotones, 
their sum $\sigma$ is an entanglement monotone as well 
\cite{Note2}. If one 
of the qubits ($Z$) is separable from the other two, then 
$\tau=0={\cal C}_{Z(XY)}\Rightarrow\sigma=0$. The converse is also 
true: If $\sigma=0$ then ${\cal C}_{Z(XY)}=0$ for some permutation 
$X,Y,Z$ of the qubits, so one qubit is separable from the other two.

Bell inequalities for three qubits are constructed from the correlator
\begin{equation}
  E(\bm{a},\bm{b},\bm{c})=\langle\psi|(\bm{a}\cdot\bm{\sigma})
  \otimes(\bm{b}\cdot\bm{\sigma})\otimes(\bm{c}\cdot\bm{\sigma})
  |\psi\rangle.\label{Eabcdef}
\end{equation}
Here $\bm{a},\bm{b},\bm{c}$ are real three-dimensional vectors of 
unit length that define a rotation of the Pauli matrices 
$\bm{\sigma}=(\sigma_{x},\sigma_{y},\sigma_{z})$. One chooses 
a pair of vectors $\bm{a},\bm{a}'$, $\bm{b},\bm{b}'$, and 
$\bm{c},\bm{c}'$ for each qubit and takes the linear combinations
\begin{eqnarray}
  {\cal E}&=&E(\bm{a},\bm{b},\bm{c}')+E(\bm{a},\bm{b}',
  \bm{c})+E(\bm{a}',\bm{b},\bm{c})-E(\bm{a}',\bm{b}',\bm{c}'),
  \nonumber\\\label{Enoprimedef}\\
  {\cal E}'&=&E(\bm{a}',\bm{b}',\bm{c})
  +E(\bm{a}',\bm{b},\bm{c}')+E(\bm{a},\bm{b}',\bm{c}')
  -E(\bm{a},\bm{b},\bm{c}).\nonumber\\
  \label{Eprimedef}
\end{eqnarray}

Mermin's inequality \cite{Mer90} reads $|{\cal E}|\leq 2$, while 
Svetlichny's inequality \cite{Sve87} is $|{\cal E}-{\cal E}'|\leq 4$. 
We calculate the two quantities
\begin{equation}
  {\cal M}_{\rm M}=\max|{\cal E}|,\;\;
  {\cal M}_{\rm S}=\max|{\cal E}-{\cal E}'|.\label{MMandSdef}
\end{equation}
The maximization is over the six unit vectors $\bm{a}$, $\bm{b}$, 
$\bm{c}$, $\bm{a}'$, $\bm{b}'$, $\bm{c}'$ for a given state 
$|\psi\rangle$. The largest possible value is reached for the 
GHZ state (${\cal M}_{\rm M}=4$ and ${\cal M}_{\rm S}=4\sqrt{2}$). 
The W state has ${\cal M}_{\rm M}=3.05$ and ${\cal M}_{\rm S}=4.35$. 
Any violation of the Svetlichny inequality implies irreducible 
tripartite entanglement \cite{Mit02}. In contrast, states in 
which one qubit is separable from the other two may still violate 
the Mermin inequality, up to ${\cal E}=2\sqrt{2}$.

The maximization over the two unit vectors $\bm{a}$, $\bm{a}'$ can 
be done separately and analytically. The maximization over the 
remaining four unit vectors was done numerically. Before showing 
results for the full 5-parameter family of states (\ref{fiveparameters}), 
it is instructive to first consider the 3-parameter subfamily
\begin{eqnarray}
  |\Phi\rangle&=&\cos\theta_{1}\left| {1\choose 0} 
  {1\choose 0} {1\choose 0}\right\rangle\nonumber\\
  &&\mbox{}+\sin\theta_{1}\left|{0\choose 1} {\cos{\theta_{2}}
  \choose \sin\theta_{2}} {\cos\theta_{3}\choose \sin\theta_{3}} 
  \right\rangle,\label{threespinors}
\end{eqnarray}
with real angles $\theta_{i}$. These states are all in the GHZ class, 
so we avoid for the moment the complication introduced by the W class. 
The physical significance of states of the form (\ref{threespinors})
 is that they are generated in optical \cite{Zei97} or electronic 
\cite{Bee03} schemes to produce three-particle entanglement from two 
independent entangled pairs. (Notice that the second and third qubit 
become separable upon tracing over the first qubit.)

\begin{figure}
  \includegraphics[width=8cm]{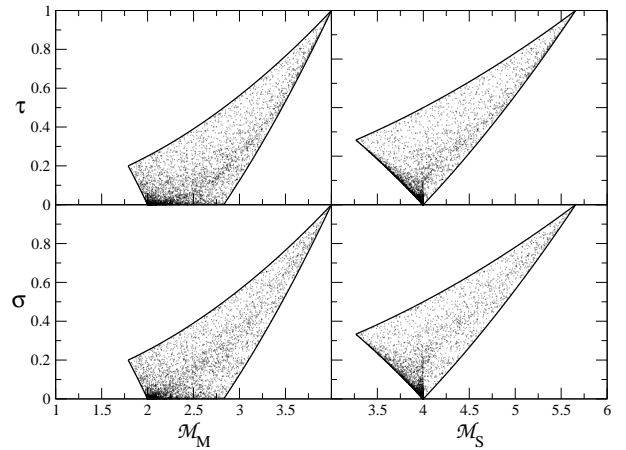}
  \caption{
  Numerically determined maximal violation of the Mermin 
  (${\cal M}_{\rm M}$) and Svetlichny (${\cal M}_{\rm S}$) inequalities
  for the three-parameter state (\protect\ref{threespinors}). A range 
  of values for the entanglement measures $\tau$ and $\sigma$ give the
  same maximal violation. The solid curves are the upper and lower 
  bounds (\protect\ref{boundM}) and (\protect\ref{boundS}). 
  \label{fig3parameter}
  }
\end{figure}

The numerical data for the 3-parameter states, shown in 
Fig.\ \ref{fig3parameter}, fills a region bounded by 
\begin{eqnarray}
  &&\max\left(1-{\textstyle\frac{1}{4}}{\cal M}^{2},0,
  {\textstyle\frac{1}{8}}{\cal M}^{2}_{\rm M}-1\right)
  \alt\tau,\sigma\alt{\textstyle\frac{1}{16}}{\cal M}^{2}_{\rm M},
  \nonumber\\
  &&\label{boundM}\\
  &&|{\textstyle\frac{1}{16}}{\cal M}^{2}_{\rm S}-1|\alt\tau,
  \sigma\alt{\textstyle\frac{1}{32}}{\cal M}^{2}_{\rm S}.
  \label{boundS}
\end{eqnarray}
These bounds on $\tau,\sigma$  do not have the status of exact 
analytical results (hence the symbol $\alt$), but they are 
reliable representations of the numerical data \cite{Note1}. 
Note that the same 
violation of the Svetlichny inequality gives a tighter lower bound 
on $\tau,\sigma$  than the Mermin inequality gives, due to the fact 
that 2--1 separable states are eliminated \cite{Mit02}.

\begin{figure}
  \includegraphics[width=8cm]{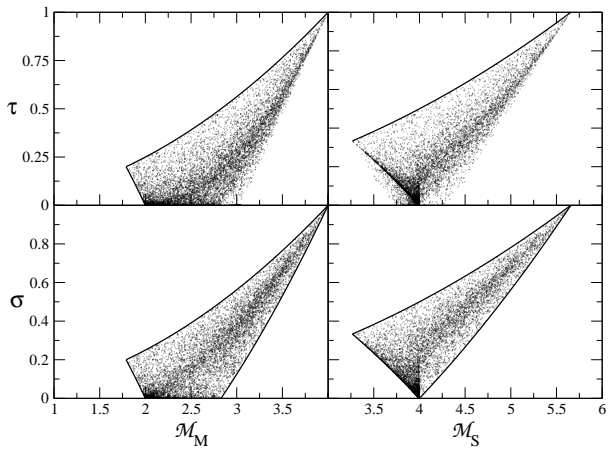}
  \caption{
  Same as Fig.\ \protect\ref{fig3parameter}, but now for the 
  general 5-parameter state (\protect\ref{fiveparameters}). 
  \label{fig5parameter}
  }
\end{figure}

For the 3-parameter states (\ref{threespinors}) in the GHZ class 
there is no advantage 
in using $\sigma$ over $\tau$. Both entanglement measures are bounded 
in the same way by the Bell inequalities. That changes when we turn to
the general 5-parameter states (\ref{fiveparameters}), which also contain
states in the W class. We see from Fig.\ \ref{fig5parameter} that
the bounds (\ref{boundM}) and (\ref{boundS}) still apply to $\sigma$. 
However, the tangle $\tau$ drops below the previous lower bound, due to 
the fact that it can not distinguish W states from separable states.

In conclusion, we have constructed an entanglement monotone $\sigma$ for
three qubits which, unlike the tangle $\tau$, can detect entanglement of both 
the GHZ and W types.  We have investigated numerically the relation 
between the entanglement measures $\sigma$, $\tau$ and the maximal 
violation of Bell inequalities (both of the Mermin and Svetlichny form).
The upper and lower bounds reported here have already been put to use 
in the design of a protocol for the detection of tripartite entanglement 
in the Fermi sea \cite{Bee03}.  Alternatively, if one wants to do 
better than  a bound, one could use the interferometric circuit 
proposed recently for the tangle \cite{Car03}, which, with a small 
modification, can be used to measure $\sigma$ as well.

We thank W. K. Wootters for drawing our attention to the merits 
of the Svetlichny inequality. This work was supported by the 
Dutch Science Foundation NWO/FOM and by the U.S. Army Research 
Office (Grant No.\ DAAD 19-02-1-0086).

\end{document}